%% file: main.tex
\documentclass[sigchi]{acmart}

\usepackage{url}
\usepackage[utf8]{inputenc}
\usepackage{xcolor}
\usepackage[normalem]{ulem}
\usepackage{makecell}

\acmYear{2020}
\copyrightyear{2020}
\acmConference[COMPASS '20]{ACM SIGCAS Conference on Computing and Sustainable Societies}{June 15--17, 2020}{, Ecuador}
\acmBooktitle{ACM SIGCAS Conference on Computing and Sustainable Societies (COMPASS '20), June 15--17, 2020, , Ecuador}
\setcopyright{acmlicensed}

\setcopyright{none}

\title{SunDown: Model-driven Per-Panel Solar Anomaly Detection for Residential Arrays}

\author{Menghong Feng} 
\affiliation{ \institution{University of Massachusetts, Amherst}}
\email{mfeng@umass.edu}

\author{Noman Bashir}
\affiliation{ \institution{University of Massachusetts, Amherst}}
\email{nbashir@umass.edu}

\author{Prashant Shenoy}
\affiliation{ \institution{University of Massachusetts, Amherst}}
\email{shenoy@cs.umass.edu}

\author{David Irwin}
\affiliation{ \institution{University of Massachusetts, Amherst}}
\email{irwin@ecs.umass.edu}

\author{Beka Kosanovic}
\affiliation{ \institution{University of Massachusetts, Amherst}}
\email{kosanovic@umass.edu}

\begin{document}

\input{abstract}
\maketitle

\input{introduction}
\input{background}

\input{solar_anomaly_detection}
\input{classifying_solar_anomalies}
\input{evaluation}

\input{related}
\input{conclusion}

\bibliographystyle{acm}
\bibliography{paper}

\end{document}

%% file: abstract.tex
\begin{abstract}
There has been significant growth in both utility-scale and residential-scale solar installations in recent years, driven by rapid technology improvements and falling prices. Unlike utility-scale solar farms that are professionally managed and maintained, smaller residential-scale installations often lack sensing and instrumentation for performance monitoring and fault detection. As a result, faults may go undetected for long periods of time, resulting in generation and revenue losses for the homeowner. In this paper, we present SunDown, a sensorless approach designed to detect per-panel faults in residential solar arrays. SunDown does not require any new sensors for its fault detection and instead uses a model-driven approach that leverages correlations between the power produced by adjacent panels to detect deviations from expected behavior. SunDown can handle  concurrent faults in multiple panels and perform anomaly classification to determine probable causes. Using two years of solar generation data from a real home and a manually generated dataset of multiple solar faults, we show that our approach has a MAPE of 2.98\% when predicting per-panel output. 
Our results also show that SunDown is able to detect and classify faults, including from snow cover, leaves and debris, and electrical failures with 99.13\% accuracy, and can detect multiple concurrent faults with 97.2\% accuracy.
\end{abstract}

%% file: introduction.tex
\section{INTRODUCTION}
\label{sec::introduction}


Recent technological advances and falling hardware price have led to significant growth in the deployment of renewable solar within the electric grid. The cost of solar deployments have dropped to less than \$2.75 per watt in recent years~\cite{seia_2019_insight} and have become competitive with traditional energy sources. As a result, utility-scale and residential-scale solar deployments have experienced sustained growth across the world, with more than 2.6GW of deployments in 2019 Q3 in the US alone~\cite{seia_2019_insight}.

Typically, larger utility-scale solar farms are professionally monitored and maintained for optimal performance---they are instrumented for monitoring real-time generation to identify production issues, and also cleaned frequently to reduce dust or pollen. Researchers have also suggested using drones carrying thermal cameras to identify and locate faults in large solar arrays \cite{arenella2017drones}. However, the majority of 
solar installations today are small-scale installations, often on residential rooftops, with capacities of less than 10 kW in 2018~\cite{seia_2018_review}. Due to cost reasons, such systems lack sensing and instrumentation that may be present in larger utility-scale solar farms. Further, monitoring of these systems is left to homeowners, who lack the technical expertise for this task.  At best, system performance may be monitored at a coarse-grain system-wide basis to determine system-level issues.
As a result, it is not uncommon for residential solar arrays to encounter power anomalies or other local faults that go undetected for long periods of time, resulting in a loss of generation and revenue for the owner. While it is possible to add sensors and instrumentation for real-time monitoring, doing so for small-scale installations increases their cost, and is challenging to do for millions of installations that are already operational without such capabilities. 

To address these challenges, in this paper, we present \emph{SunDown}, a sensor-less approach for detecting generation faults in small-scale solar arrays on a per-panel basis (the terms fault and anomaly are used interchangeably in this paper).  Our approach assumes that per-panel generation information is available from the array---an assumption that holds true for any installation that uses micro-inverters or DC power optimizers---and uses a model-driven approach to detect when the panel output deviates in an anomalous manner from the model-predicted output. Our approach is based on machine learning and can detect physical anomalies such as snow obstructions, leaves, and electric faults at panels. Our approach seeks to identify and alert solar owners of such issues in a timely manner so that they can be rectified to avoid production losses. 

In designing, implementing, and evaluating our SunDown system we make the following contributions. 

\begin{itemize}
\item We present a model-driven approach, based on machine learning, that leverages correlations in the generated output between adjacent panels to predict the expected output of a particular panel and flags anomalies when the model predictions deviate from the expected values. Unlike prior work that has performed system-level fault detection, our approach is designed to perform more fine-grain fault detection at a per-panel level. Further, our approach can handle and detect multiple concurrent faults in the system, a key challenge that has not been addressed by prior work.
\item We present a random forest-based classification technique to classify the probable cause of the observed fault. To validate our approach, we construct two labelled datasets of solar anomalies: a two year dataset from a real-home with real snow cover anomalies that we hand label using ground truth information, and a solar anomaly dataset that we construct with a twenty-panel array by injecting synthetic faults such as dust, leaves, and open circuit faults. Since there is a dearth of solar anomaly datasets, we release both datasets and our code as open-source tools to the community.
\item We conduct a detailed experimental evaluation of our methods. We show that our approach has a MAPE of 2.98\% when predicting per-panel output, which shows the efficacy of using neighboring panels to perform model-driven predictions. Our results also show that SunDown is able to detect and classify faults such as snow cover, leaves, and electrical failures with 99.13\% accuracy for single faults and is able to handle concurrent faults in multiple panels with 97.2\% accuracy.
\end{itemize}






%% file: background.tex
\section{BACKGROUND}
\label{sec::background}
In this section, we present background on residential solar arrays and solar anomaly detection. 

\begin{figure}[t]
\centering
\includegraphics[width=0.3\textwidth]{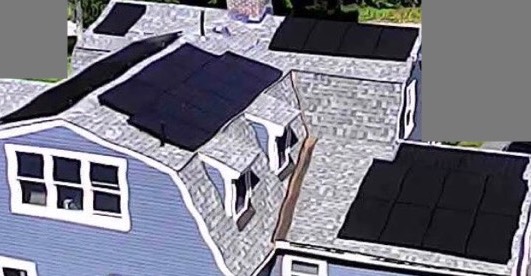}\\
\includegraphics[width=0.3\textwidth]{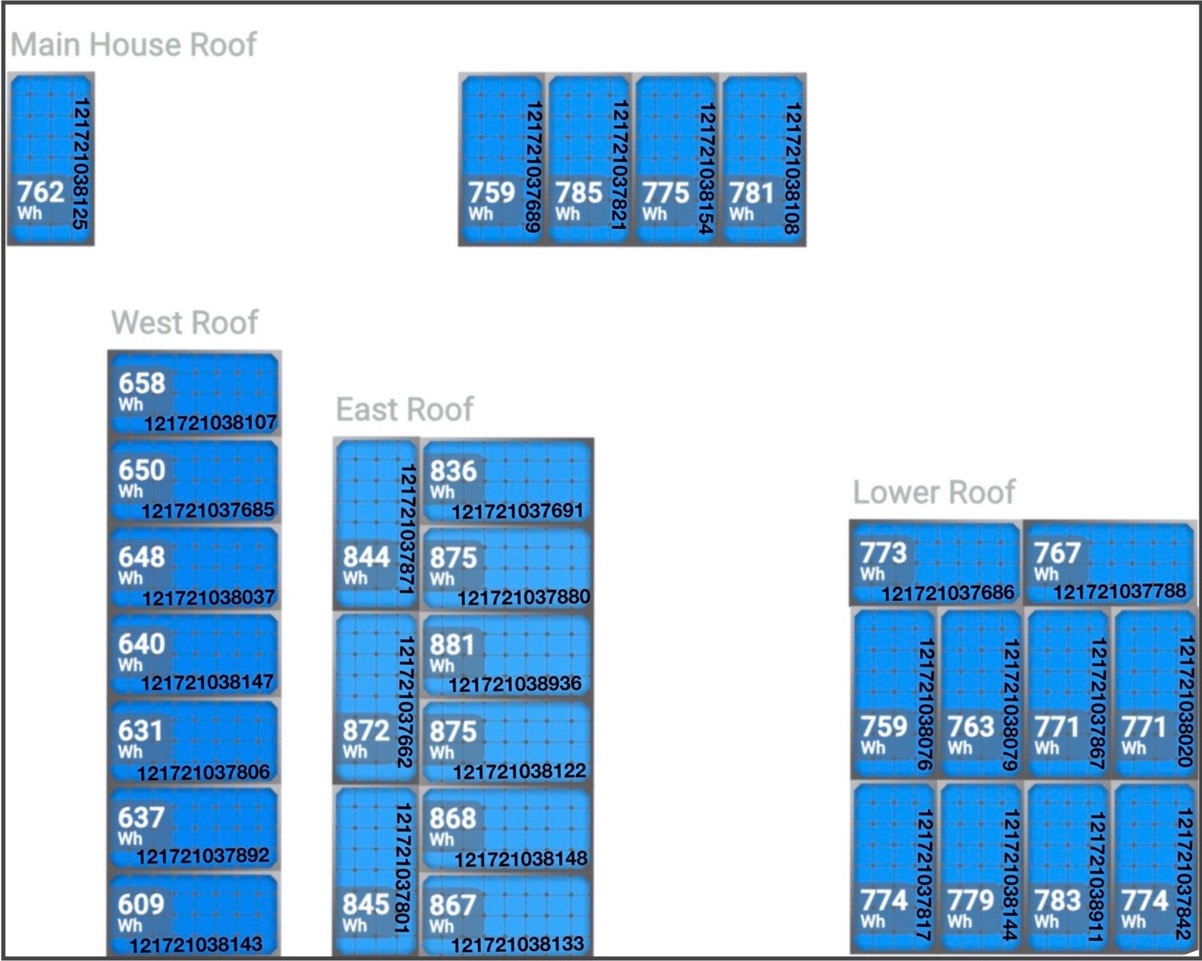}
\caption{A residential solar array (top) with 31 panels deployed on four roof planes, and real-time panel-level generation data from the array (bottom)  }
\label{fig:david_house}
\end{figure} 

\subsection{Residential Solar Arrays}
\label{sec::back::rooftop}

Our work primarily focuses on residential solar arrays, such as ones often found on residential rooftops. Such installations are typically small-scale installations with capacities of 10kW or less and comprise a few to a few dozen solar panels (see Figure \ref{fig:david_house}). Since we are interested in monitoring anomalies and faults at a per-panel level, we assume that the power generation of the array can be monitored at a per panel level. 

This is a reasonable assumption in practice since many residential arrays are equipped with micro-inverters (e.g. Enphase micro-inverters \cite{enphase}) or DC power optimizers \cite{solarEdge-power-opt} on each panel that are designed to track and independently optimize the power generation of each individual panel. Such installations, which are now commonplace, are advantageous since they maximize the total system output even for deployments that span multiple roof surfaces and under partial shading-effects. 
As shown in Figure \ref{fig:david_house}, such systems provide real-time per-panel generation data, which is essential for our approach.
Other than knowledge of per-panel output, we do not assume any other sensors or instrumentation on the residential solar installation. Thus, we seek to develop a sensor-less approach for per-panel solar anomaly detection.

\subsection{Solar Generation}
\label{sec:back:generation}

It is well-known that solar generation at any site depends directly on the amount of sunlight -- solar irradiance -- received at that location. The solar irradiance is a function of the latitude and longitude of that location and the season of the year \cite{iyengar2017cloud}. Of course, the weather---specifically cloud cover---can reduce the solar irradiance at a particular site.  

For the purpose of this work, we assume that per-panel solar generation on any given day can be reduced to two factors: \emph{transient}, which comprises of factors that temporarily impact power output, and \emph{faults} which comprise of factors that have a prolonged negative impact on output.  

Transient factors include weather conditions such as cloud cover, wet panels caused by rain or dew, as well as site specific factors such as shading caused by nearby trees or other structures. We can classify transient factors into two classes---common or local. Common transient factors are those that impact all panels of a particular site such as overcast condition or rain. Local factors are those that impact a particular panel, or a group of panels, but not all of the panels at that site. For example, many shading effects may impact a portion of the site, depending on the foliage and the location of the sun. 

\subsection{Solar Faults}
\label{sec::back::anomaly}
Anomalies (also referred to as faults) in our case are defined to be factors that cause a persistent drop in production but can be rectified by the owner of the site. We are particularly interested in the following three types of faults (1) snow cover on one or more panels, (2) partial occlusions such as bird droppings, dust or leaves on a panel, (3) electric faults such as module failure, short circuits or open circuits. These faults cause either a reduction in output or zero output for a particular panel or a subset of panels. 

Due to their close proximity to one another, multiple panels in a residential array may experience the same fault---for example, snow may cover multiple adjacent panels (or even the entire system), resulting in concurrent faults.
Of course, a site may also suffer a full system outage, which is also a fault but is easier to detect than those that cause partial outages or partial output reduction.

\subsection{Problem Statement}
\label{sec::back::ps}
Consider a solar array with $N$ solar panels. We assume that the panels are mounted on a residential roof and may be mounted on one or multiple roof planes. Note that in the latter case, panels will have different tilts and orientations. We assume that the power generated by each panel can be monitored in real-time and that the weather at the site is also known (e.g. from a weather service). Given such a setup, our problem is to design a technique that monitors the power output of each panel and the entire system, and labels the observed output in each time interval (e.g. a day) as normal or abnormal. Further, our technique should identify specific solar panels in the system that are experiencing faults and also determine possible cause of the fault (e.g. snow, partial occlusion, or electric fault).

%% file: solar_anomaly_detection.tex
\section{PER-PANEL SOLAR ANOMALY DETECTION}
In this section, we describe our model-driven approach for per-panel solar fault detection and how we can build on this approach to perform multiple fault detection. We first describe the basic idea, followed by the details of our models and algorithms.

\subsection{Basic Idea}
Consider a solar installation with $N$ panels. Suppose that $k$ panels are experiencing an anomaly that result in a reduction, or loss, of output from those panels. Initially, let us assume $k=1$ (only one panel out of $N$ is faulty). Later on, we show how our approach can be extended to handle multiple, concurrent faults where $k>1$.  

Since all $N$ panels are mounted on the same roof in close proximity of each other, it follows that they experience highly correlated weather conditions, and produce similar output. Thus, our "sensorless" approach first constructs a model to predict the expected output of a panel from $n$ neighboring panels ($n\geq 1$). For example, a simple predictor is one that uses the mean output of $n$ neighboring panels to estimate a particular panel's output. Under normal conditions, since adjacent panel outputs are highly correlated, the model prediction will match the observed output of that panel with high accuracy. Note that any $n$ out of the available $N$ panels can be chosen to model the output of a particular panel. A useful heuristic is to use the ``closest'' $n$ panels to the one being predicted or to use the $n$ panels on the same roof plane since they will have higher correlations than those on a different roof surface of the same house. In our evaluation, we experimentally evaluate the accuracy of these heuristics and also evaluate the value of $n$ that yields sufficient accuracy. 

When a panel experiences an anomaly, however, the model predictions will continue to estimate the "normal case" output of that panel, while the observed output will deviate from this normal case.  
If the deviation is "large" and persists over an extended period of time, it is indicative of a fault, rather than an error in the model prediction. The cause of the fault can be separately determined by analyzing amount of loss or the power pattern exhibited by the panel. Such a model-driven approach only uses the observed output of panels to detect anomalies---no other instruments or sensors are needed for anomaly detection unlike some other approaches \cite{arenella2017drones}. 

\subsection{Model-Based Predictions}
Based on the above intuition, we now present two model-driven techniques for predicting the power output of an individual panel using neighboring panels. Our first model is based on linear regression and uses only power output of panels as input parameters to make predictions. Our second model is  based an a probabilistic graphical model and half-sibling regression. 

\subsubsection{Linear Regression-Based Model}
\label{sec:anomaly:regression}
Since the power generated by solar panels in close proximity of one another are highly correlated, we can use regression to predict the output of a panel given the observed output of neighboring panels.

Let $P_{i}$ denote the observed power output of panel $i$ at time instant $i$. Let us assume we wish to predict the output of panel $i$ using $n$ other panels. Typically we can choose $n$ nearest panels, or $n$ panels on the same roof plane, out of the $N$ total panels on the roof. A linear regression model  allows us to estimate the output of desired panel as  a linear function of the others:

\begin{align}
P_i = w_i P_{i1} + w_2 P_{i2} + w_3 P_{i3} + ... + w_n P_{in} + \epsilon_i
\end{align}
where $X=\{i_1, i_2, ..., i_n\}$ is the set of $n$ panels used to model the output of the $i^{th}$ panel. We can use linear regression to estimate the weight $w_i$ that minimize the error term $\epsilon_i$.

Such an approach yields $N$ distinct regression models, one for each panel in the system, where each model makes prediction using the observed output of $n$ other panels. To determine if a panel has a fault, we compare the model predictions at time $t$, $P_i(t)$
with the observed value $\hat{P}$. If the difference between the model predictions and observed values is large and persists over a period of time (e.g., a day or multiple days), the approach flags that panel as faulty.

\subsubsection{Graphical Model and Half-Sibling Regression}

Our second model is based on a recently proposed machine learning technique called half-sibling regression that uses a Bayesian approach to remove the effects of confounding variables \cite{scholkopf2016modeling}. This approach has been used by astronomers to remove noise from measurements of multiple telescopes observing the same phenomena. The main intuition behind the approach can be understood from the astronomy use-case. Suppose that $n+1$ telescopes are observing the same object such as star. The observations will have some ``common'' noise introduced by factors such as air pollution or haze that impact visibility of the object. Furthermore, each telescope will have local factors such as instrument calibration error that introduce additional local errors. If we use observations of $n$ telescopes to estimate the expected observation of the $(n+1)$-st instrument, and take the difference between the observed and predicted values, we are left with the local errors (``anomalies'') at that instrument. In our case, we have $n+1$ solar panels ``observing'' the sun---their power output represent their observations of the sun. All panels see common factors such as clouds that introduce similar output reductions in the power values. Further, each panel has local factors such as shade (transient factors) or faults that can result in additional reductions in the power output. If we use $n$ panels to predict the output of the $n+1$-st panel using a Bayesian model, the difference between the predictions and observed output should isolate local factors including the effect of faults.  This is the intuition behind using the Bayesian approach of \cite{scholkopf2016modeling}.   

\begin{table}[t]
    \centering
    \begin{tabular}{|c|c|c|}
        \hline
         & SolarClique & SunDown\\
        \hline
        Per-Panel faults & No & Yes\\
        \hline
        System-wide faults  & Yes & Yes\\
        \hline
        Multiple faults & No & Yes\\
        \hline
        \makecell{Anomalies \\
        Detected} 
        & \makecell{System-wide \\
        electrical} 
        & \makecell{Snow, electrical,\\
        occlusion}\\
        \hline
    \end{tabular}
    \caption{A comparison of the state-of-the-art SolarClique system and our SunDown approach}
    \label{approach_diff}
\end{table}

More recently, this approach was used in a system called SolarClique\cite{iyengar2018solarclique} to predict the output of an entire array using nearby solar arrays. We draw inspiration from the  half-sibling regression paper \cite{scholkopf2016modeling} and SolarClique \cite{iyengar2018solarclique} for SunDown's anomaly detection, but point out important differences between the SolarClique method and our approach as shown in table \ref{approach_diff}. First, SolarClique is designed for system-level predictions (predicting the total generation of an entire array) and does not have the capability of making fine-grain per-panel predictions, which is the focus of our method. Second, a key technical limitation of SolarClique is that it assumes a single fault can occur at a time, and that the system is not capable of scenarios where multiple arrays are faulty.  This is a reasonable assumption for SolarClique since it uses $n$ arrays from $n$ different homes to predict the output of a specific home, and  faults across arrays and homes can be assumed to occur independently. In our case, since panels are in close proximity to one another, the same fault (e.g., snow) can impact multiple panels, and faults therefore no longer occur independently. Since the independence assumption of SolarClique does not hold in our case, a key technical improvement over prior work is our ability to handle multiple faults (as discussed in the next section). For simplicity, we first assume a single fault in the entire system and present our approach. We then relax the assumption in the next section and show how the basic model can be extended to handle multiple concurrent faults. A final difference is that SolarClique did not focus on fault classification (and only detects large system-level electrical failures) while SunDown can identify
multiple types of faults, including snow cover, occlusion faults and electrical faults.

\begin{figure}[t]
  \centering
  \includegraphics[width=0.7\linewidth]{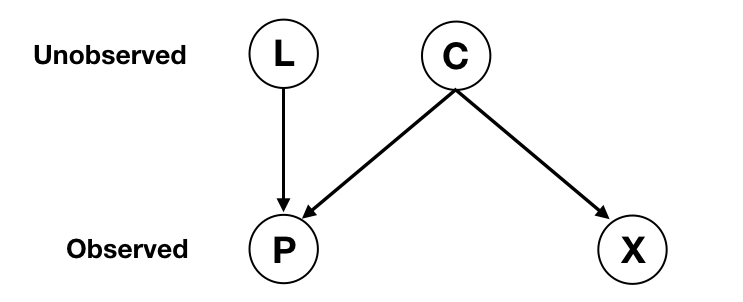}
  \caption{Graphical model representation}
  \label{graphical_model}
\end{figure}

To describe our Bayesian model, let $P$ be a random variable denoting the power output of a particular panel. Let $X$ denote a random variable representing the power output of $n$ other panels in the system. Hence, $X$ is a vector of size $n$. Let $C$ denote the confounding variables that impact both $X$ and $P$. In our case, $C$ denotes common confounding variables such as cloud cover that have the "same" impact on panels. Let $L$ denote the local factors that impact the output of a panel. $L$ will include transient factors, including partial shading, as anomalies that locally impact $P$. The relationship between $P$, $X$, $L$, and $C$ can be captured using a (causal) graphical model as shown in  Figure \ref{graphical_model}. Since the output of each panel can be directly monitored, $P$ and $X$ are observed variables, while $C$ and $L$ are latent unobserved variables.

As can be seen, $P$ depends on both $L$ and $C$ while $X$ depends only on $C$ (and is independent of $L$). $C$ impacts $X$, and when conditioned on $P$, $P$ becomes a "collider", making $X$ and $L$ dependent. To reconstruct $L$ using half-sibling regression, we assume the following additive model
\begin{align}
P = L + f(C)
\label{eq_1}
\end{align}

Since $C$ is unobserved, we can use $X$ (which is observed) to approximate $f(C)$. If $X$ exactly approximate the function $f(C)$, we can then compute $f(C)$ on $E[f(C)|X]$. Even otherwise, if $X$ is a sufficiently large vector, it can yield a ground approximation. Thus, we can use $X$ to predict $P$ and recover $L$ from Equation \ref{eq_1} as 

\begin{align}
\hat{L} = P - E[P|X]
\label{eq_2}
\end{align}
Note that $\hat{L}$ estimates both anomalies and transient factors, and the impact of transient factors must be removed from $L$ to estimate the anomaly.

Given these concepts, our algorithm to estimate the amount of production loss due to anomalies is as follows:

We first use regression to estimate $P$ using $X$. This is similar to the linear regression method from the prior section. The regression yields $E[P|X]$ - an estimate of $P$ given the observed output of $n$ neighboring panels that constitute $X$. Since $P$ itself is observed, subtracting $E[P|X]$ from $P$ yields an estimate of the output loss $\hat{L}$ due to transient factor and anomalies as shown in Equation \ref{eq_2}. A key difference between linear regression model of section \ref{sec:anomaly:regression} and here is that we use bootstrapping to construct multiple regression model by subsampling the data (instead of a single regression model) and use an ensemble method based on Random Forest that uses the mean of multiple models to estimate $E[P|X]$.

Next, since $\hat{L}$ contains effects of transient factors such as shade on panels as well anomalies, we must remove the impact of transient factors to obtain the "true" anomalies. We can use time series decomposition to extract the seasonal component that represents shading effect that occur daily at set time periods and remove it from $\hat{L}$~\cite{iyengar2018solarclique}. The remainder of $\hat{L}$ represents production loss at that panel due to any anomalies.

Under normal operation $\hat{L}$ will be close to zero (no anomalies and no loss of output). When $\hat{L}$ is significant and persistent over a period of time, our model-driven approach flags an anomaly in the panel.

\subsection{Handling Multiple Concurrent Faults}
Both our regression and Bayesian models use the power output of $n$ panels to predict the expected output of another panel. A very important assumption is that the $n$ panels being used as inputs to the model are non-faulty and hence be used to predict the normal case output of another panel. An anomaly is flagged when the model prediction of normal case output deviates from the observed output, indicating the presence of an anomaly. 

This approach works well when there is only one faulty panel in the system - which implicitly implies that all remaining panels are non-faulty and any model that uses some of these remaining panels to make predictions will have ``clean''  non-faulty inputs.
 However, due to the close proximity of panels, anomalies such as snow cover, dust, leaves, are likely to impact multiple panels. In this case, some of the inputs to the model may come from faulty panels, causing model prediction to have high errors. 

Of course, if $n$ is made large and only a small number of panels are faulty, the model may be able to tolerate the "noise" in a small number of inputs and still produce reasonable accurate prediction. However, many residential rooftops may have a small number of panels, which means $n$ can not always be large. Hence, we need an explicit
method to tolerate the impact of multiple concurrent faults in the system.

Observe that our models use {\em any} $n$ out of $N$ total panels to predict the output of panel $i$. Thus, it is possible to construct {\em multiple models for each panel} by choosing different subsets of $n$ panels out of $N$, and then using them as inputs to predict the output of panel $i$. In the normal case (no faults), all of these models show similar predictions for panel $i$'s output. However, when multiple panels are faulty, any model that uses faulty panels as input will have higher errors while a model that uses all non-faulty inputs will continue to provide good predictions. Our goal then is to construct multiple models for each panel using our Bayesian or regression method, and then choose one of these models at each instant that uses non-faulty inputs.

To do so, we need to distinguish  between faulty and non-faulty inputs. 
However, since the models are themselves being used to detect faults, we need a different method to determine which inputs are possibly faulty. To do so, we  use a
solar forecasting approach that predicts the output of the solar panel based on weather forecasts. There is extensive work on solar forecasting using weather forecasts and any
such model can serve our purpose. We use a machine learning forecasting-based model that uses the location of the system (longitude and latitude), time of day, past power observations  and near-term weather forecasts (e.g., sunny, cloudy) to estimate the output of a panel \cite{iyengar2017cloud}.  This model, and many others, have been implemented into the Solar-TK open-source library~\cite{solar-tk}, which we leverage to design a custom forecasting model for each panel in the system using near-term future weather forecasts.

Suppose that $P_i(t)$ is the estimate of power output of a panel $i$ based on this forecasting model. If $P_i(t) - \hat{P}_i$ is large, it implies that expected output differs from the prediction and the panel is possibly a "noisy" input.  Our per-panel forecasting models  perform these prediction for each panels and labels it as "noisy input" or "normal input". Any model that uses one or more noisy panel  as an input should be eliminated from consideration for anomaly detection purposes.  

That is, SunDown chooses any regression or Bayesian model (out of multiple models for a panel constructed from different subsets comprising $n$ panels) such that all inputs to that model are labelled normal. 

\begin{figure}[t]
\centering
  \includegraphics[width=0.4\textwidth]{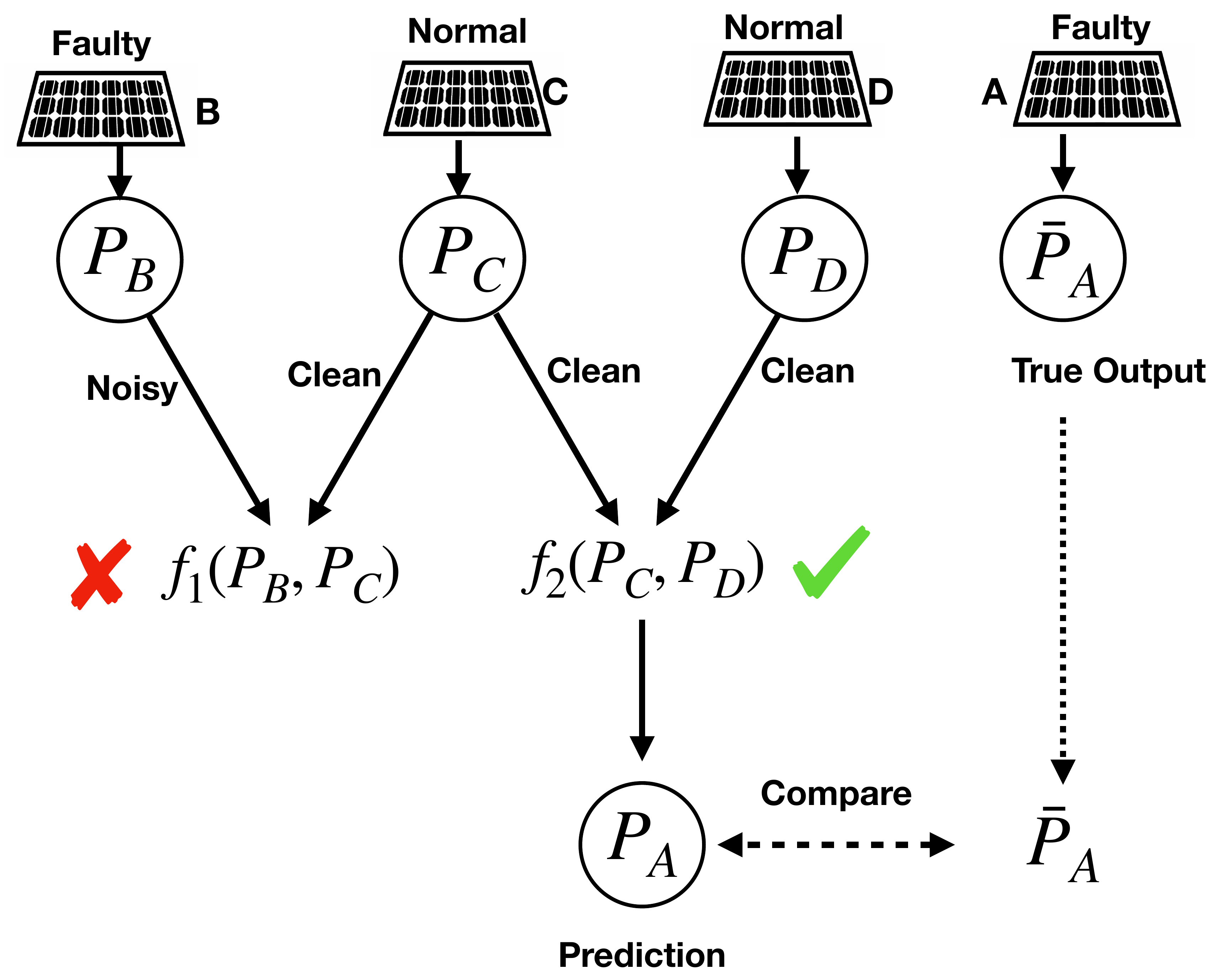}
  \caption{A forecasting model is used to ensure non-noisy inputs to our Bayesian model.}
  \label{abcd_process}
\end{figure}

Consider the following example to illustrate the process (figure \ref{abcd_process}). Suppose that a solar rooftop install has 4 panels: $A$, $B$, $C$, $D$. We wish to predict the output of panel $A$ using two other panels. Suppose both $A$ and $B$ are faulty. Let us assume we have the following two half-sibling regression-based Bayesian models, $f_1$ and $f_2$ to predict $P_A$, the power output of panel $A$
\begin{align}
P_A = f_1(P_B, P_C)
\label{model_1}
\end{align}
\begin{align}
P_A = f_2(P_C, P_D)
\label{model_2}
\end{align}
where model  $f_1$ predicts $A$ using panels $B$ and $C$ as inputs, while $f_2$ predicts $A$ using $C$ and $D$. Our approach first predicts $P_A$, $P_B$, $P_C$, and $P_D$ using per-panel machine learning solar forecasting models for each of the four panels ~\cite{solar-tk}. Since $A$ and $B$ are faulty, they get labeled noisy inputs. Hence, $f_1$ is eliminated from consideration since one of its inputs, $P_B$, is a noisy input and $f_2$ is chosen for prediction since both its inputs, $P_C$ and $P_D$, are  labelled "normal". Using model $f_2$  yields a better estimate for $P_A$ than model $f_2$. Note that, doing so enables us to handle concurrent faults--we can avoid using faulty panels as model inputs, and at the same time, use our Bayesian method to identify the presence of multiple faults. 

Note that although our solar forecasting models also provide an estimate of the panel's output, they are not suitable for anomaly detection. This is because they use weather forecasts of cloud cover, along with other parameters, to estimate a panel's output. Forecasts of future weather are inherently error-prone, which means the the forecasting model will also have higher errors. Using the solar forecasting model directly for anomaly detection will have higher false positive (due to model errors). In contrast, the Bayesian approach uses actual power output observations to estimate a panel's output for purposes of anomaly detection, which yields a more accurate model and reduces changes of false positives. This is the reason we use forecasting models to only identify noisy inputs; incorrectly labeling a panel as noisy due to forecasting error only causes some of the models to suppressed for anomaly detection, and does not impact accuracy of the remaining models for finding faulty panels.


\begin{figure*}[t]
\centering
\begin{tabular}{ccc}
\includegraphics[width=0.3\textwidth]{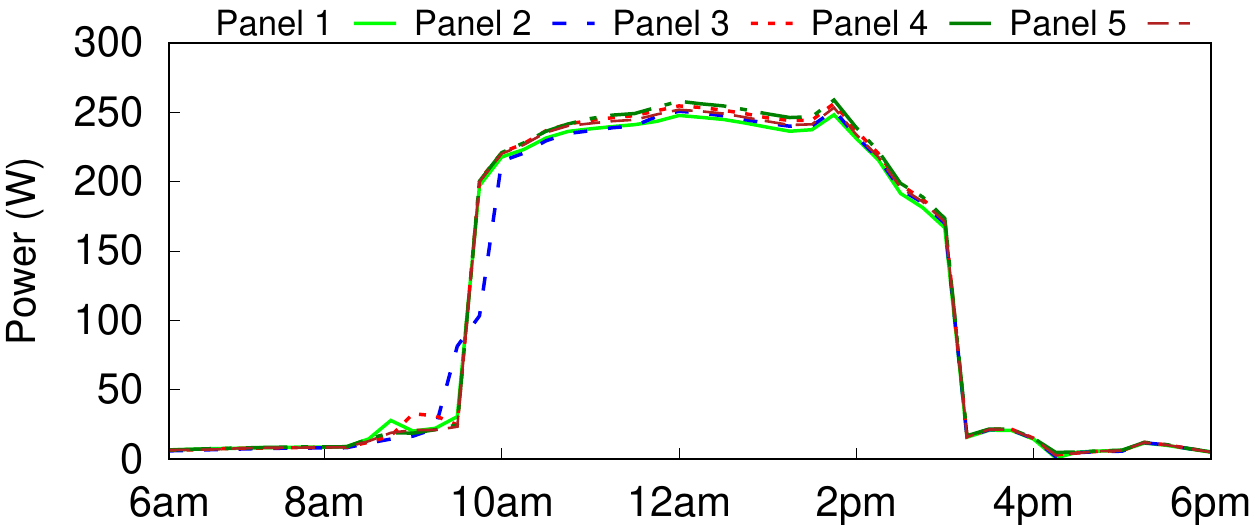} &
\includegraphics[width=0.3\textwidth]{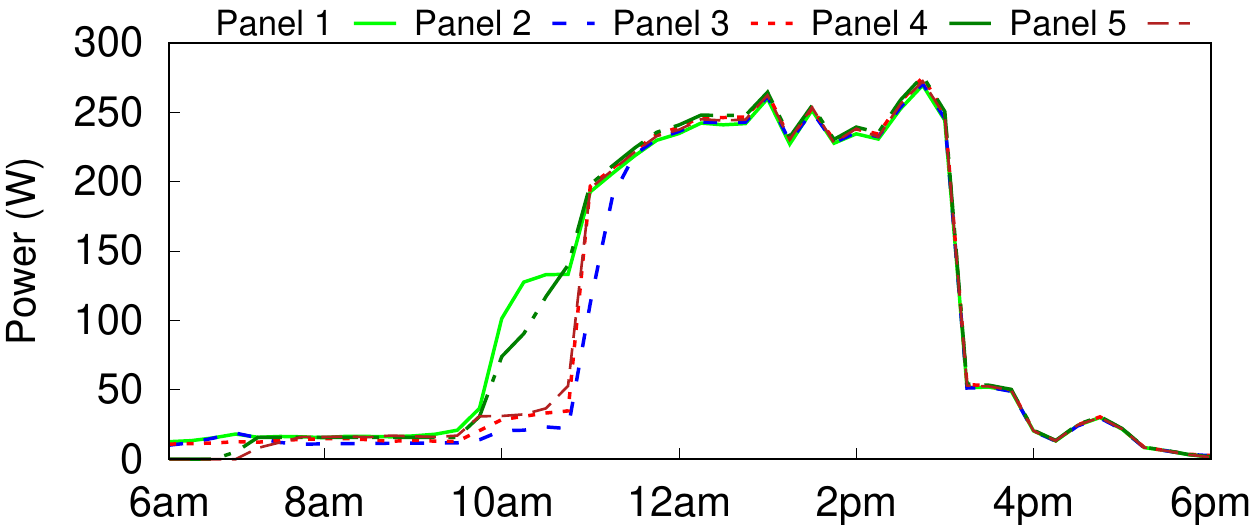} &
\includegraphics[width=0.3\textwidth]{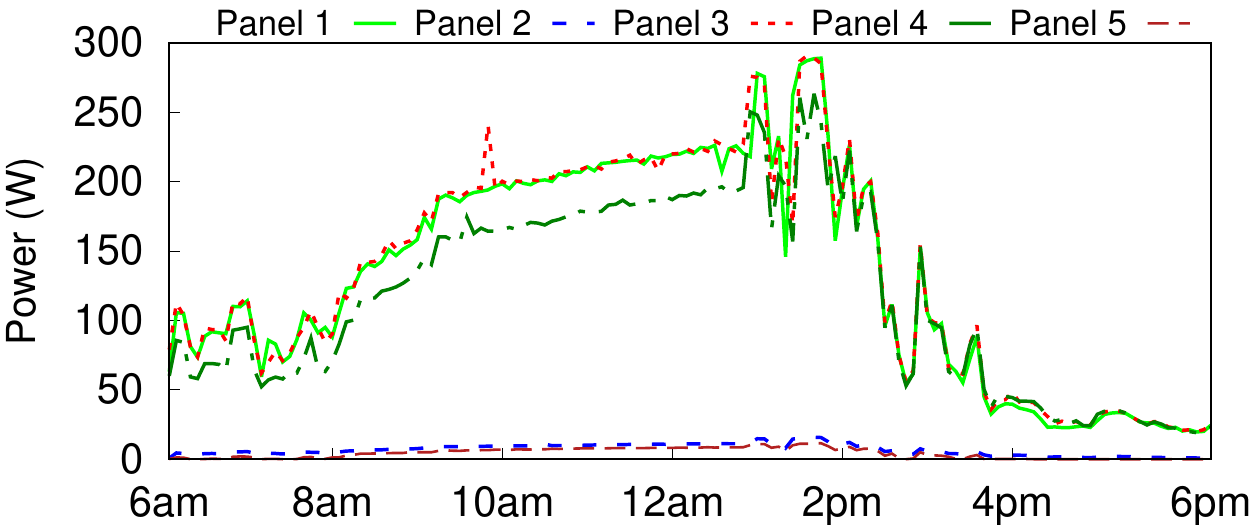}\\
(a) Normal & (b) Partial Shading & (c) Snow Fault\\ 
\end{tabular}
\caption{Residential home power output on an example day under (a) normal condition, (b) partial shading on some panels on east side, (c) snow covering on some of the panels.}
\label{fig:dataset}
\end{figure*}

%% file: classifying_solar_anomalies.tex
\section{CLASSIFYING SOLAR ANOMALIES}

While the previous section presented model-driven approaches to detect the presence of anomalies in one or more panels, in this section, we present a classification approach to determine the possible causes of the output loss seen at the panel(s).

\begin{figure}[t]
\centering
\includegraphics[width=0.45\textwidth]{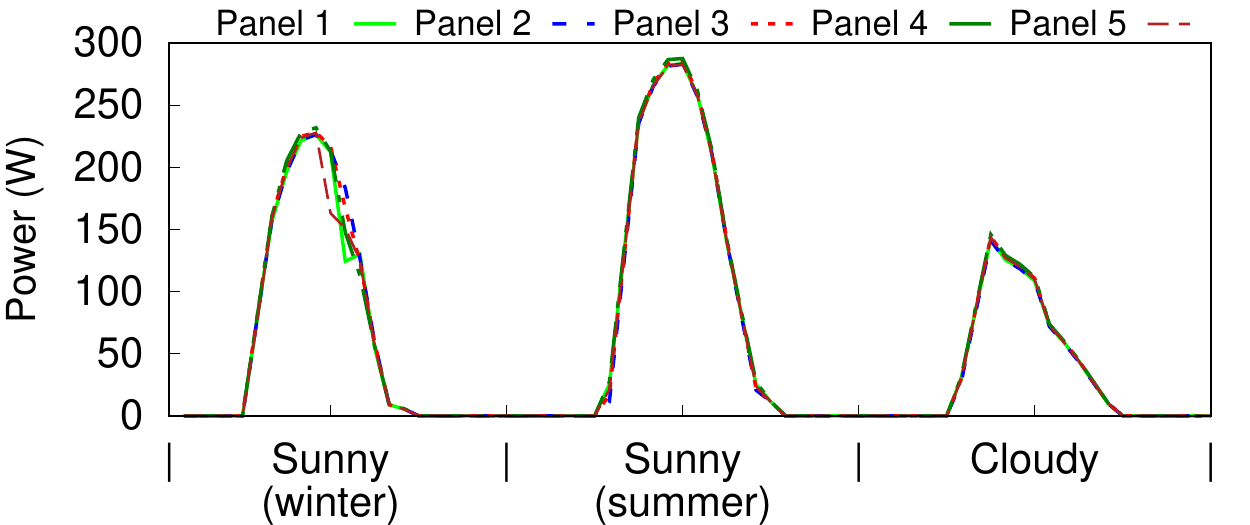}
\caption{Panel output on sunny day in summer and winter and a cloud day.}
\label{fig:seasonal}
\end{figure}

\subsection{Solar Anomaly Open Dataset}
To assign a possible cause to an observed output loss, we must analyze the observed power pattern and match it to the "power signature" exhibited by different type of solar faults. However, this requires that we have ground truth data for various type of faults, which is challenging since there are no open datasets of solar faults available for research use (solar farm operators likely have such data but have not released it to others). Consequently, we need to gather our own data with ground truth information on solar faults.

Our anomaly dataset contains data from two residential scale solar installations:
\begin{enumerate}
  \item a 31-panel, 9kW solar installation (Figure \ref{fig:david_house} top) that experienced multiple snow cover anomalies (Figure \ref{fig:david_house} bottom) over its two year lifetime
  \item a 20-panel ground mounted solar installation (Figure \ref{leaf_cover}) where we systematically introduce  anomalies such as dust, leaves, electrical faults,  etc., to mimic real-world faults and measure its impact on the output.
\end{enumerate}
We discuss each dataset in more detail before describing our classification method.

\subsubsection{Snow Anomaly Dataset}
This dataset comes from a residential solar array deployed on a home in Northern America (location details removed for double blind renewing). The house contains 31 rooftop panels, mounted on four different roof planes, as shown in figure \ref{fig:david_house}(bottom). Each panel is a 320W LG panel with an Enphase micro-inverter that can optimize the panel's output independently of the rest. As noted earlier, micro-inverters optimize and report panel-level generation data, which is a prerequisite for our models.

We have been gathering data from this system for over two years and have per panel generation information at 5 minute granularity from September 2017 to February 2020. We have also gathered weather data for the location from Darksky and NOAA weather service. 

The only real anomaly encountered by this system over the two year period is snow cover, following a snow storm (the area receives frequent snowfall in the winter). Depending on how long the snow sticks on the panels following a snow event, snow-covered panels may produce little or no output. As snow melts, some panels generate output, while others stay covered with snow (Figure \ref{fig:dataset}(c)).

\begin{figure}[t]
\centering
\includegraphics[width=0.4\textwidth]{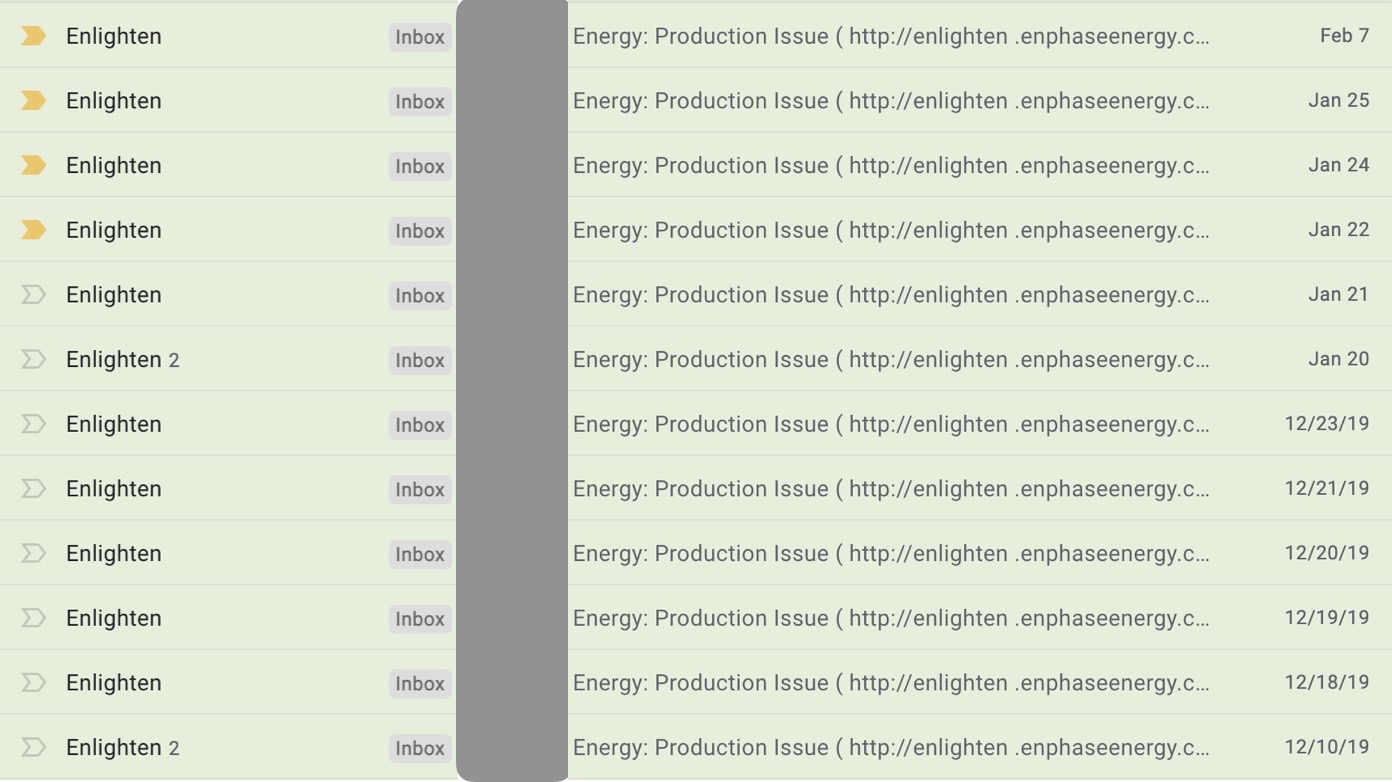}
\caption{Snow event email alert}
\label{email_alert}
\end{figure}

We have two sources of ground truth to label snow faults. First, the Enphase system sends an email to the homeowner when it observes near zero output for an entire day, as shown in figure \ref{email_alert}. The email indicates a "possible production" issue at the system.
Second, Darksky and NOAA provide  past weather data, such as snow events and the extent of the snowfall at an location. 

\begin{figure}[t]
\centering
\includegraphics[width=0.4\textwidth]{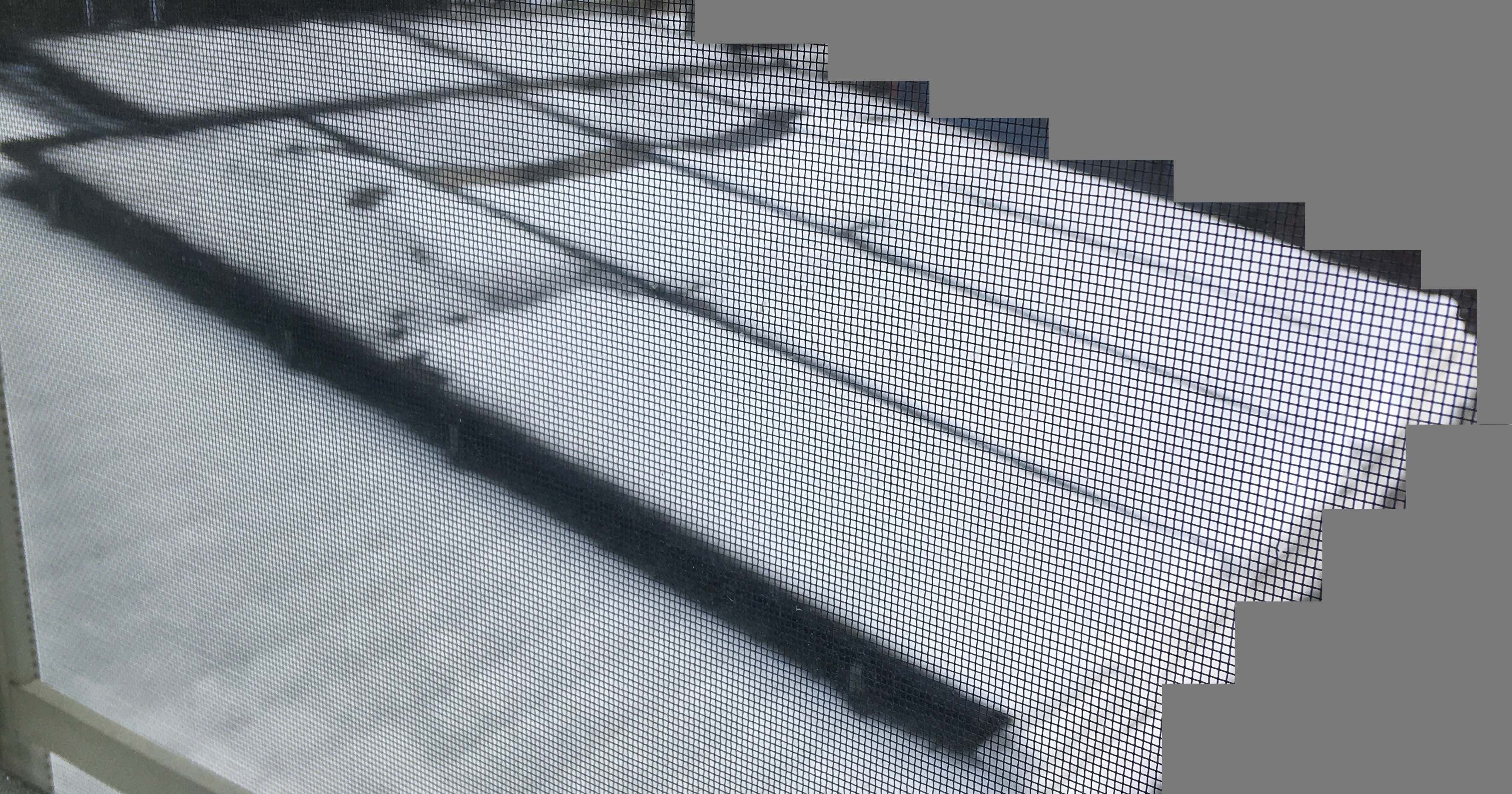}
\caption{Lower roof under snow}\label{roof_snow}
\end{figure}

We use both sources of information (which match closely with each other) to manually inspect the per panel generation data on a snow day and the following several days. We then hand label each panel's output as normal (if it produces any output) or as a snow anomaly (if the panel output is near zero). This yields a hand-labelled dataset of snow anomalies.


\subsubsection{Solar Anomaly Dataset}

\begin{figure}
\centering
  \includegraphics[width=0.4\textwidth]{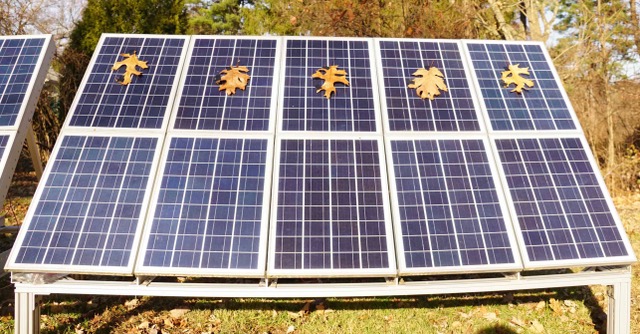}
  \caption{A synthetic leaf occlusion fault in our experimental array.}
  \label{leaf_cover}
\end{figure}

Using our 20-panel ground mounted experimental array and sensors to measure its power output,  we carefully introduced several types of anomalies onto specific panels, and measured its impact on the power output. 
We conduct several data gathering experiments over a period of several weeks under different conditions (sunny, partially overcast, overcast etc) and gathered data for the following anomalies.
\begin{enumerate}
  \item Leaf occlusion:  We introduced different number leaves on  panels (partial occlusion anomaly) and measured its impact
  \item Dust occlusion: We added different amounts of dust on the panels and measured its impact
  \item Water drops  occlusion: We  add varying amount of water drops on the panel and measure its impact. This is designed to mimic morning dew on panels, which is not a true anomaly but a weather effect
  \item Open circuit fault: We  used a variable potentiometer to introduce a high resistance seen by the panel to mimic an open circuit fault and mesured its impact.
\end{enumerate}
This hand-crafted anomaly dataset, along with photographs and labels, provides an additional source of data for our experiments. For example, Figure \ref{leaf_cover} shows leaves on the panel that emulate  a partial occlusion fault. 
Figure \ref{fig:dataset}(a) and (c) depicts the output of the panels in normal conditions and under a snow fault, respectively. 
Figure \ref{fig:synthetic-data} (a) and (b)  illustrate the power output under synthetically-generated open circuit fault and a partial occlusion fault. 
We have released both datasets to the reseearch community.

\subsection{Classifying Anomalies}

Given anomalies detected by our Bayesian model
we use a random forest classifier to label the possible cause of the fault for each panel that is faulty.  The classifier needs to distinguish between three types of faults: snow, partial occlusion and open circuit.  Note that partial snow over on a panel and partial occlusion faults both result in diminished, but non-zero output. Full snow cover on a panel and open circuit faults both yield zero output. To distinguish between these cases, we first sample 40 randomly chosen points over an entire day and compute the percentage reduction in power output when compared to the model predictions for each of these points. This power loss vector is a key feature to our classifier. We also use two other features: month of the year and snow depth values from NOAA weather service. We train our random forest classifier using a training dataset of real snow and synthetic anomalies.  Depending on the season (winter versus other seasons) and the observed power loss over a period of time, our classifier can label the probable cause of fault for each panel. Our approach can also label system-wide faults, caused either by a system-wide electrical failure or full snow cover on the entire system, both of which cause near total loss of power output.

%% file: evaluation.tex
\section{EXPERIMENTAL EVALUATION}
\label{sec:eval}


We evaluate SunDown by quantifying (1) the accuracy of model-based power inference where we infer the output of a single panel using nearby panels, (2) the impact of parameters such as number of panels, roof geometry, and weather, and (3) the accuracy of our anomaly classification. We quantify the accuracy of predicting a  panel's output using Mean Absolute Percentage Error (MAPE) between the inferred output and the actual solar generation, as below. 
\begin{align}
MAPE = \frac{1}{m} \sum_{t=1}^{m}\Big| \frac{P_O(t)-P_I(t)}{\bar{P_O}} \Big|
\label{mape_eq}
\end{align}
where $m$ is the number of samples, $P_O(t)$ is the observed solar power at time $t$, $P_I(t)$ is the inferred power at time $t$, and $\bar{P_O}$ is the mean of observed power generation. Equation \ref{mape_eq} is an alternative form of standard MAPE where we replace the denominator comprising a single observed value by the mean of all observed values. The alternative form avoid divide by zero issues when the denominator (and observed value) are zero.

For the anomaly detection and classification tasks, our goal is to correctly classify all the different anomalies. We use three different metrics to quantify different aspects of the classification task: accuracy, sensitivity, and specificity.
The accuracy is computed by dividing the number of correctly classified anomalies by the total number of anomalies. Sensitivity and specificity metrics are used for the unbalanced data case where the number of one category is smaller than other. The different metrics are computed as below. 

\begin{align}
Accuracy = \frac{TP+TN}{N}
\label{eq_acc}
\end{align}

\begin{align}
Sensitivity = \frac{TP}{TP+FN}
\label{eq_sensi}
\end{align}

\begin{align}
Specificity = \frac{TN}{TN+FP}
\label{eq_speci}
\end{align}
where $N$ is the total number of instances, $TP$ is the number of anomalies correctly classified, $TN$ is the number of normal days correctly classified, $FP$ is the number of normal days classified as anomalies, and $FN$ is the number of anomalies misclassified as normal days. Accuracy is used to evaluate the overall model's performance, while sensitivity and specificity are used to test how accurate the model is to correctly detect the anomalies and normal cases.  

\begin{figure}[t]
  \includegraphics[width=0.9\linewidth]{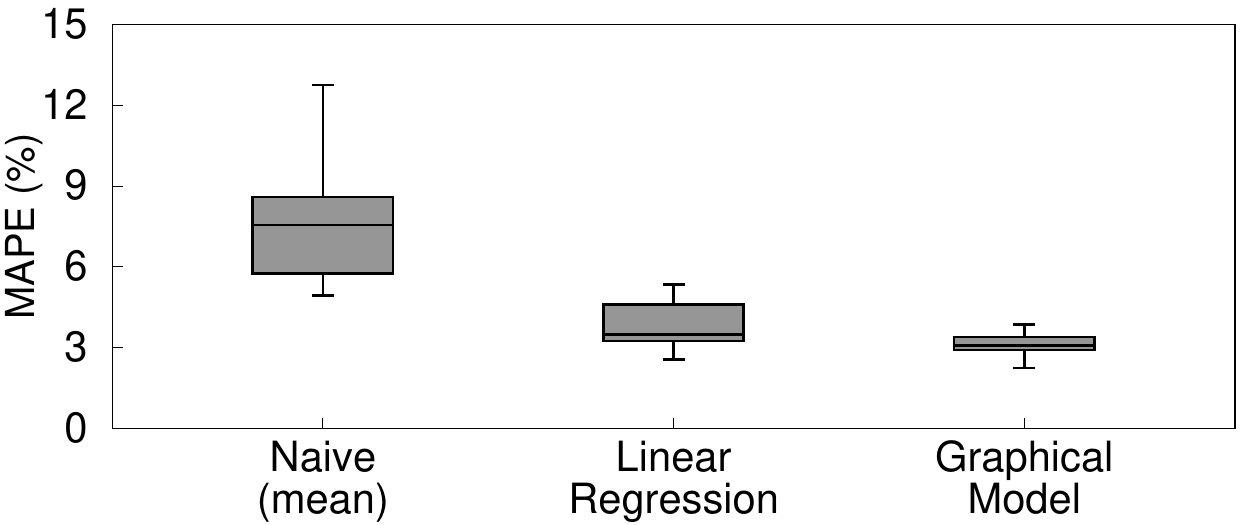}
  \caption{Machine Learning Model}
  \label{model_impact}
\end{figure}

\subsection{Prediction Model Accuracy}
We begin by evaluating the accuracy of predicting the power output of an individual panel using neighboring panels. 

\subsubsection{Machine Learning Model}
\label{sec:eval:model}
To evaluate the accuracy of model inference, we choose a test data only from the days where the site experiences no anomaly. 
We then use the normal days of the home dataset to train our linear regression and graphical model.  
We also compare their performance with a naive approach that infers the power output of a panel as the mean output of $n$ other panels. We then compare the model predictions using a test dataset and compute the MAPE values for each approach.
As shown in Figure \ref{model_impact}, the MAPE values for Bayesian model, linear regression, and naive approach are 3\%, 4\%, and 8.6\%, respectively. 
The naive approach has the worst accuracy since it  all panels produce similar output, which is not true in many cases due to panel level variations.
Linear regression  works well when the output of different panels are highly correlated and have a linear relation between them, which is not true when some of the panels experience partially shading. Our graphical ensemble learning approach is able to model non-linear relationships and  yields highest accuracy and a tight confidence interval. We use the graphical model for the subsequent experiments, unless stated otherwise. 

\begin{figure}[t]
  \includegraphics[width=0.9\linewidth]{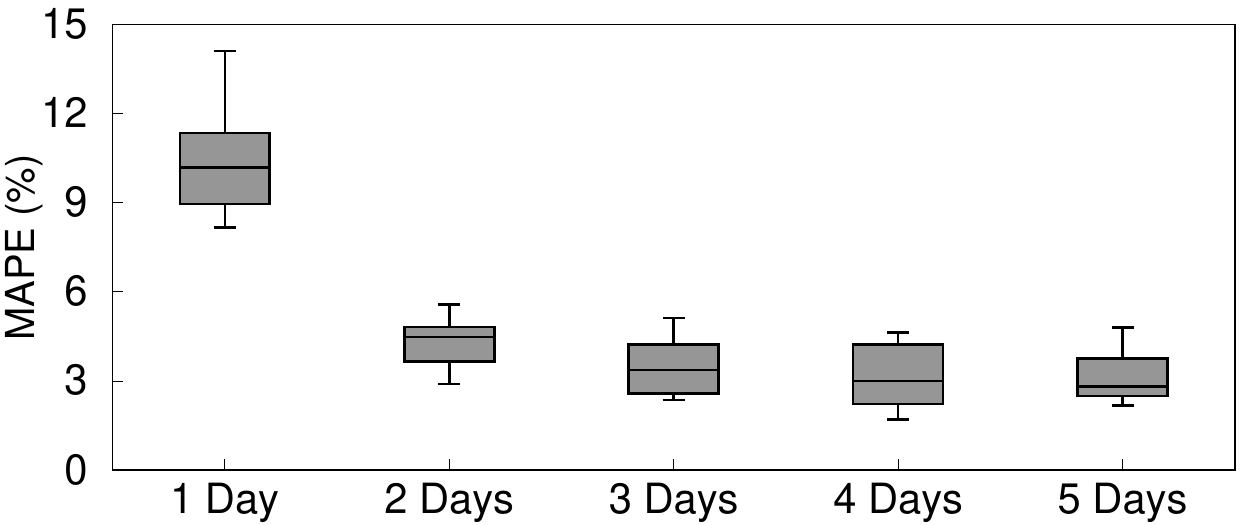}
  \caption{Size of training data required}
  \label{size_impact}
\end{figure}

\subsubsection{Impact of Training Data Size }
\label{sec:eval:training}
Next, we evaluate model accuracy for different amounts of training data.
If a model requires a lot of training data for good accuracy, it can hinder its use for solar sites that have been recently deployed or for the sites where long-term panel level data is not available. We vary the training data size (by randomly choosing a certain number of days) and  evaluate its accuracy for predicting output using a test dataset.  Figure \ref{size_impact} demonstrates that our model can achieve a decent accuracy and a 10\% MAPE with only one day of per panel data. If the number of days is increased to 4, the MAPE drops to 3.5\% and stays almost constant beyond four days. 

\begin{figure*}[t]
\centering
\begin{tabular}{ccc}
\includegraphics[width=0.3\textwidth]{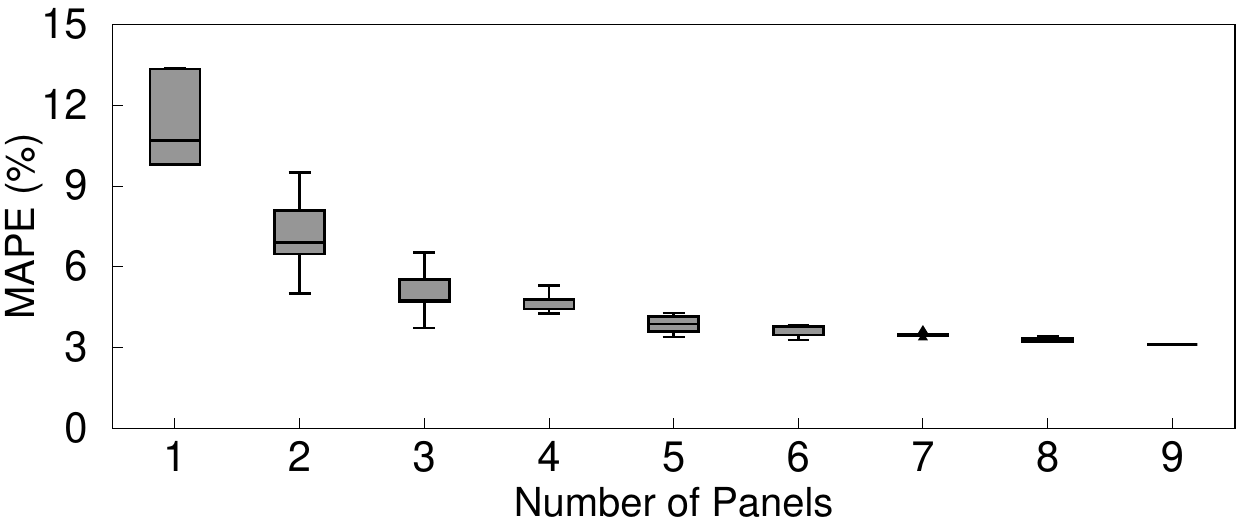} & 
\includegraphics[width=0.3\textwidth]{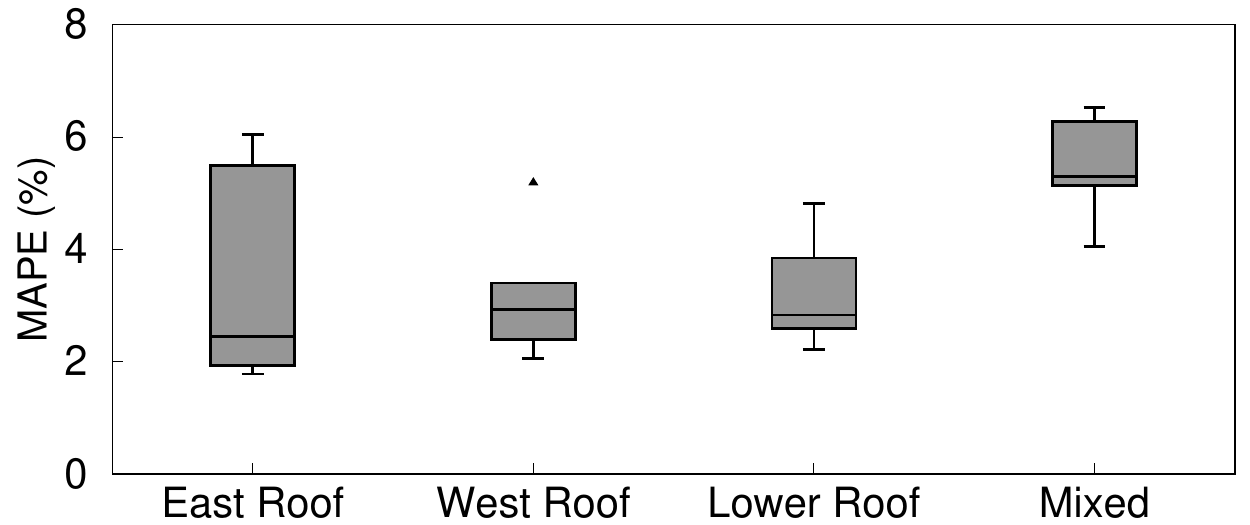} &
\includegraphics[width=0.3\textwidth]{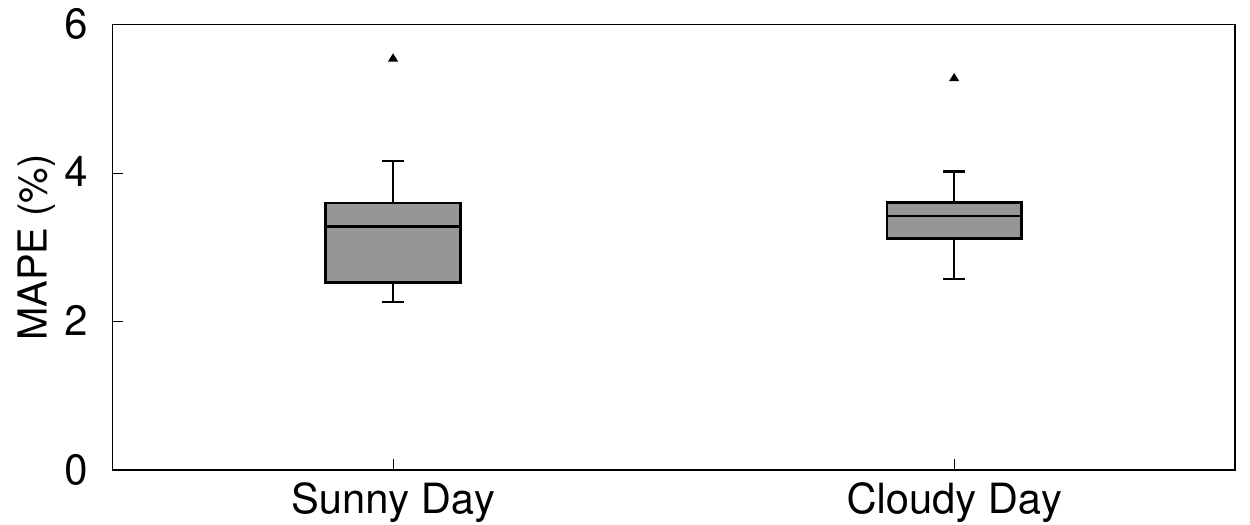}\\
(a) No. of Panels & (b) Roof Geometry & (c) Weather
\end{tabular}
\caption{Effect of various factors on the model accuracy (a) number of panels, (b)  roof geometry, and (c) weather. }
\label{fig:model-accuracy}
\end{figure*}

\begin{figure*}[t]
\centering
\begin{tabular}{ccc}
\includegraphics[width=0.3\textwidth]{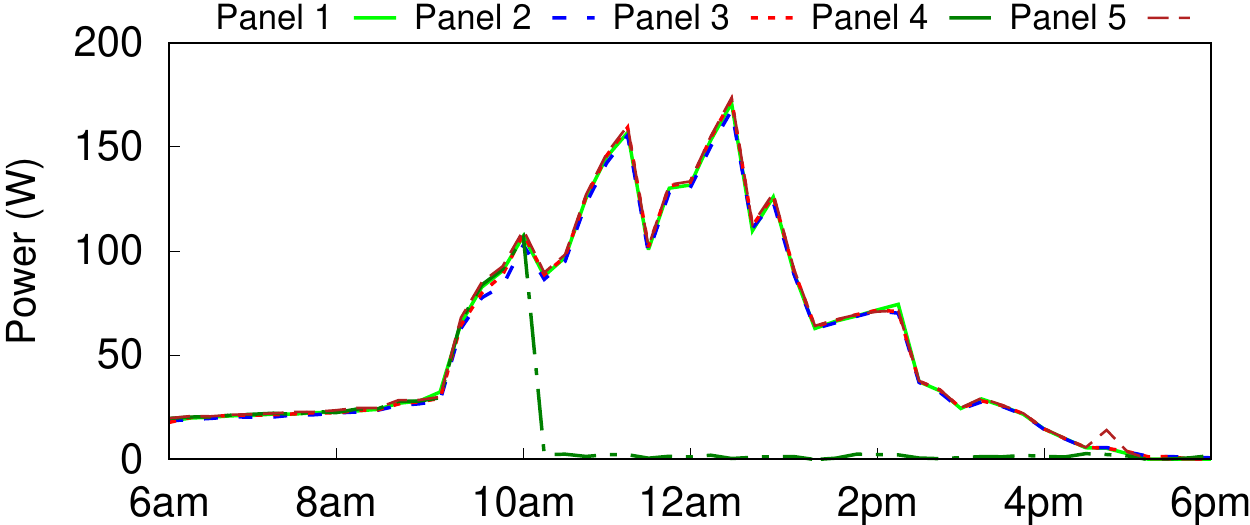} &
\includegraphics[width=0.3\textwidth]{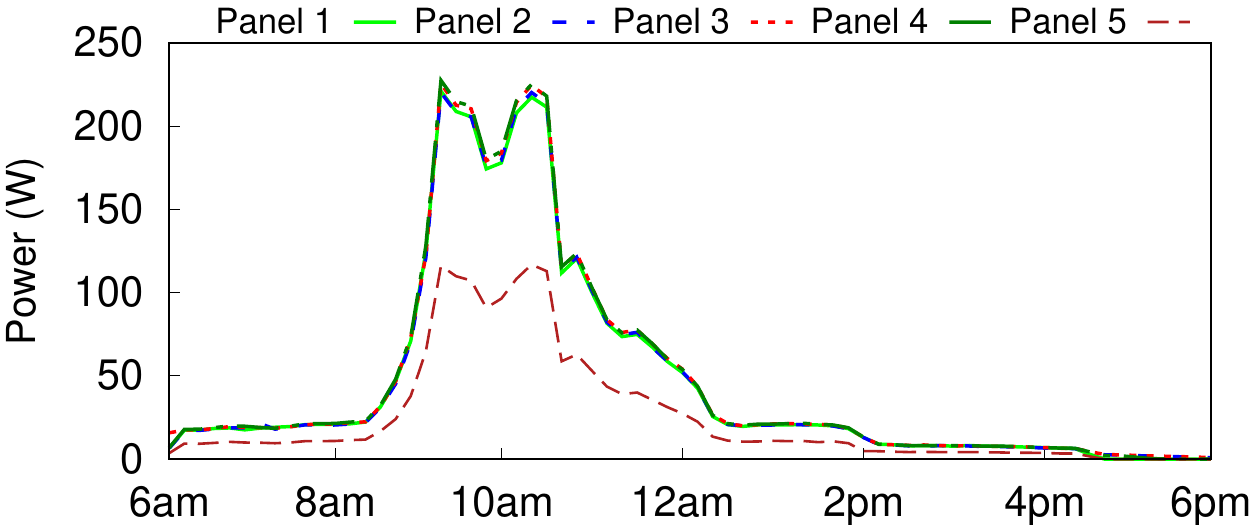} &
\includegraphics[width=0.3\textwidth]{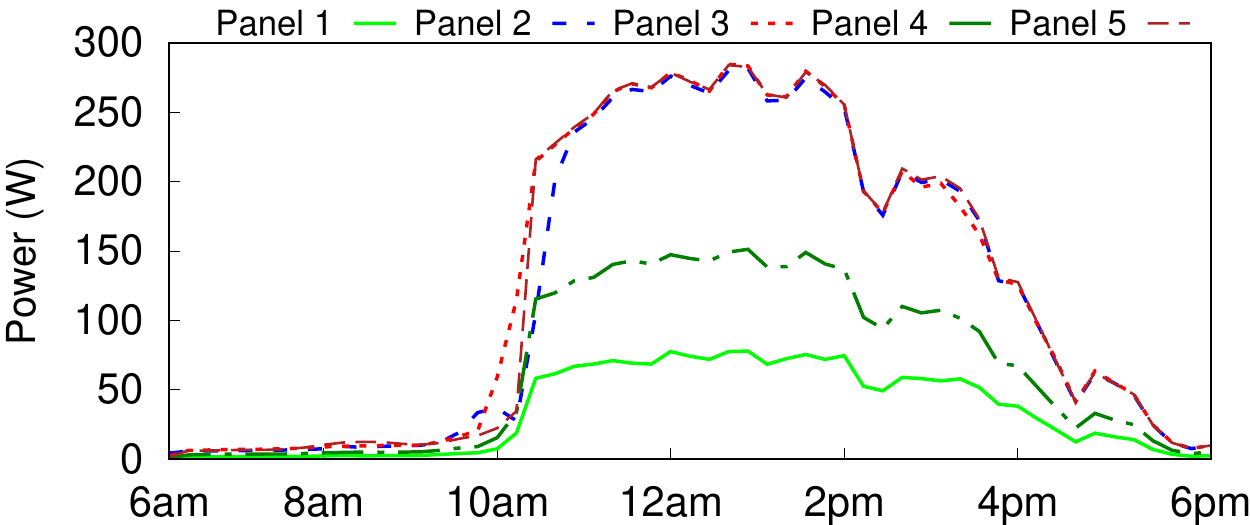} \\

(a) Synthetic Open Circuit & (b) Synthetic Object Covering & (c) Synthetic Multiple Faults\\
\end{tabular}
\caption{Synthetic fault injection with (a) open circuit fault, (b) leaves covering fault, (c) multiple leaves covering faults }
\label{fig:synthetic-data}
\end{figure*}

\begin{figure*}[t]
\centering
\begin{tabular}{ccc}
\includegraphics[width=1.6in]{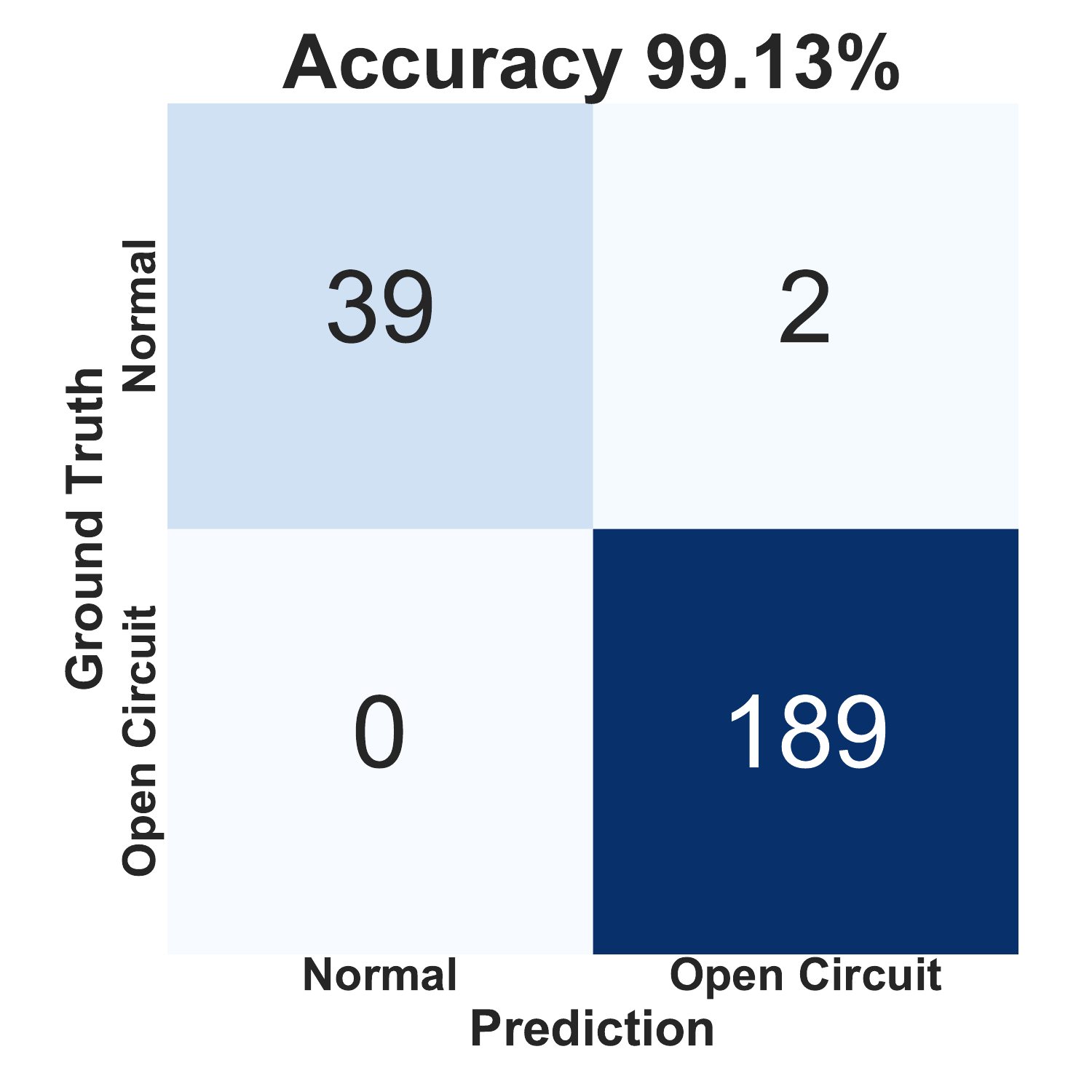} &
\includegraphics[width=1.6in]{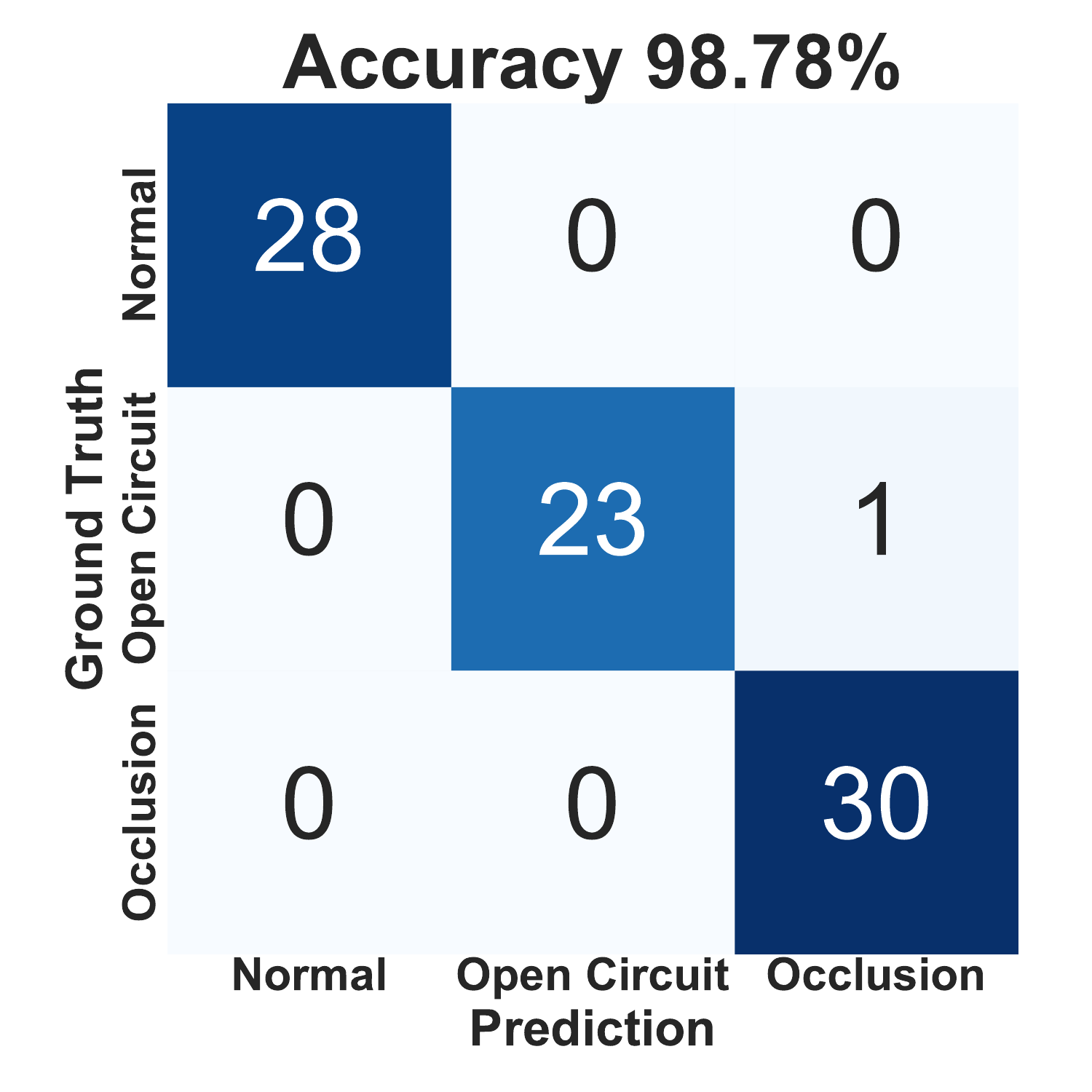} &
\includegraphics[width=1.6in]{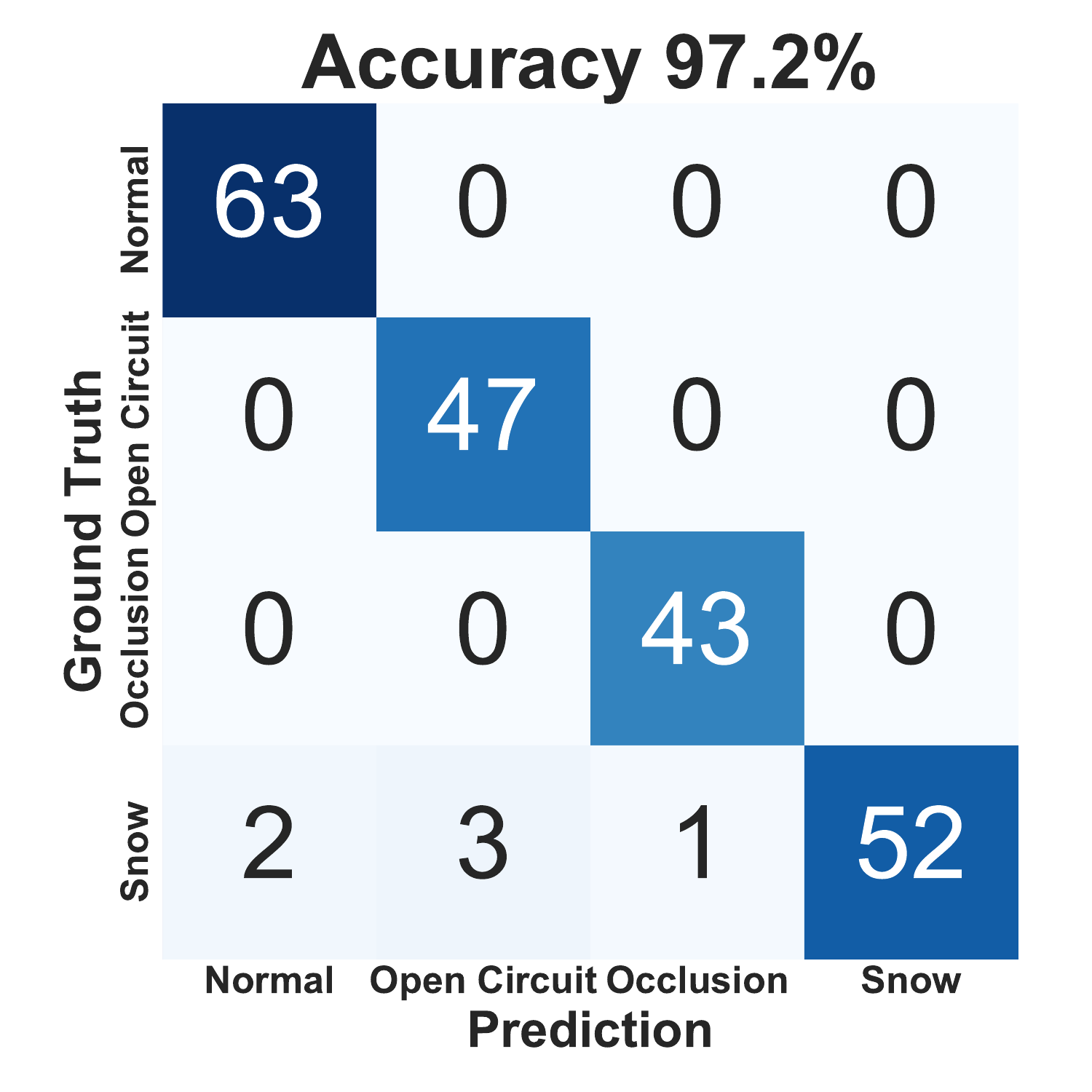} \\
(a) System level fault  & (b) Single panel fault& (c) Concurrent faults\\
\end{tabular}
\caption{Classification accuracy for (a) system-wide snow faults, (b)  single panel faults, (c) multiple panel faults.}
\label{fig:anomaly_class}
\end{figure*}




{\em Results:} Our graphical model can predict per-panel output with 2.98\% MAPE and outperforms  linear regression and a naive averaging approach. 
The random forest-based ensemble graphical model does a better job of capturing non-linear relationships among less correlated data than linear regression.  While model accuracy increases with training data size, even only four days of training data yield good accuracy.

\subsection{Impact of parameters}
We next investigate various factors that impact the inference accuracy, including  number of panels, geometry of the solar deployment and weather.

\subsubsection{Impact of Number of Panels }
\label{sec:eval:panel_num}


The individual solar panels at a site can demonstrate subtle variations in their solar output, despite their close proximity, due to panel-level dust, different tilt and orientation angles, and panel level physical faults such as cracked glass. 
To evaluate how many panels are need by a model to provide adequate accuracy,
 we vary $n$ (the number of panels used by the model as input) and compute MAPE for different $n$.
Figure~\ref{fig:model-accuracy} shows inaccuracy is  high when using less than 3 panels for inference. The accuracy improves as number of panels is increased to 5 and shows diminishing gains beyond that. The model has an average MAPE value of only 3-4\% and a very tight bound, when using 5 panels, as compared to 9\% MAPE with single panel. This result suggests that SunDown requires as little as 5 panels to be highly accurate. 


\subsubsection{Roof Geometry Impact}
\label{sec:eval:geo}


The output of a solar panel depends upon its tilt and orientation, among other factors~\cite{sundance}. Since a residential array may be installed on multiple roof planes, 
it is preferable to use panels on the same roof plane to predict others (since they will
have similar tilt and oritentation and will exhibit higher correlations).

To evalaute the effect of roof geometry, we split the home dataset into four sub-dataset based on the four roof planes whete panels are deployed. We create four graphical models to predict the power output of $i_{th}$ panel by using $n=7$ panels as inputs. For east roof, west roof, and lower roof cases, all 7 input panels are mounted side by side on the same roof plane facing the same direction. In the forth scenarios, a mixed dataset is created by combined 2 panels from each east roof and west roof datasets, and 3 panels from lower roof dataset. Figure~\ref{fig:model-accuracy}b illustrates the inference accuracy as the geometry of panels used for inference is varied. For the same roof plane, the model is highly accurate and the MAPE value is between 3\% to 3.2\%. The large variation for the east roof is due to the partial shading on some of the panels on the roof, leading to inaccurate inferences. The average MAPE of 5.5\% for the mixed dataset demonstrates that our model produces a decent accuracy even when input panels are chosen from different roof planes.
Thus, when knowledge of the roof geometry is available, it should be exploited, but the model works well even for systems where the roof geometry may be unknown causing the model to use panels from different roof planes for inference.

\subsubsection{Impact of Weather}
\label{sec:eval:weather}
The weather at a solar site, primarily cloud cover, impacts the power generation of a site. On a sunny day, all the solar panels produce similar amount of power.
However, on a cloud day, scattered clouds may only cover one or few of the panels leading to power variation across panels, which can complicate inference. Figure~\ref{fig:model-accuracy}c illustrates the effect of weather on the accuracy of the inference task. Our model achieves similar mean accuracy on both sunny and cloudy days, indicating it performs well regardless of weather. The higher variance in MAPE on a sunny day is due to shading from nearby structures, that has a more prominent impact on a sunny day over a cloudy one.



{\em Results:} Our experiments  show that the number of panels used for prediction as well as the roof geometry play an important role in the model's performance. 
We find that model yields higher accuracy when five or more panels are used for predictions and when these panels are co-located on the same roof plane. 
The weather conditions, however, do not impact model accuracy.

\subsection{Anomaly Classification Accuracy}
The previous section evaluated the accuracy of our model in predicting the output of a panel using nearby panels. We next evaluate the accuracy of model-drives approach and the classifier in detecting anomalies and classifying anomalies, respectively.  The common anomalies we consider include snow fault, open circuit, and partial occlusions due to leaves. Although, others factors such as partial shading also results in the loss of energy, we do not consider shade to be an anomaly since it it is a transient phenomena and does not need corrective action.


Our home dataset already includes real snow faults that are labelled and we evaluate the accuracy of our classifier on identifying these snow faults. We then use the synthetic faults from our solar anomaly datatset and synthetically inject them into the home data set by introducing synthetic single panel faults as well as concurrent fault and evaluate the accuracy of our classifier.
Figure~\ref{fig:synthetic-data} presents per-panel data for a typical day when electric fault or object covering anomaly has been injected into one or many panels. 



\subsubsection{Snow Fault Detection}

We first evaluate the ability of our classifier in detecting snow faults in the home dataset (recall that the data set is labelled as normal or snow for each panel). 
We extract the features from daily power output, which include Pearson's correlation coefficient, ratio of maximum observed power and the nominal panel capacity, and weather data such as snow and cloud cover and use them as inputs to our random forest classifier. Figure~\ref{fig:anomaly_class}(a) shows the confusion matrix of our classifier and shows high accuracy. Table \ref{class_metrics} shows that our approach is able identify  system-level snow faults an accuracy of 99.13\%, sensitivity of 100\%, and specificity of 95.12\%.
We note that snow faults seen in our dataset tends to be system-wide faults,  where all panels get covered with snow after a snow event and exhibit a snow fault concurrently. While it is certainly possible for only some panels to have snow cover (e.g., if snow melts unevenly across panels), our dataset presently does not have such faults.


\subsubsection{Single and Concurrent Fault Classification}

Since all observed snow faults in our dataset were system-faults, we next show that our approach is still capable of fine-grain anomaly detection and classification of a single fault and it is also capable of detecting concurrent faults in a subset of the panels.



To do so, we use our solar anomaly dataset and choose the partial occlusion and open circuit anomaly from the dataset and inject these faults into a single, randomly chosen, panel of the array; different panels have faults injected into them on different days. We use our model to detect the presence of the fault and our random forest classifier to identify the type of fault.  We next inject multiple concurrent faults of all types (snow, occlusion, open circuit) into the  array using a similar methodology and attempt to detect and classify each fault using our model and classifier (note that we need to use our concurrent fault detection approach in this case).

Figure~\ref{fig:anomaly_class}b and ~\ref{fig:anomaly_class}c
show the confusion matrix of classifying single and concurrent faults in the array.  Table \ref{class_metrics} 
shows that our model can classify single fault with accuracy of 98.78\%, specificity of 97\%, and sensitivity of 100\%. For multiple concurrent faults, the model obtains accuracy of 97.2\%, specificity of 97.06\%, and sensitivity of 97.26\%.

\begin{table}[t]
    \centering
    \begin{tabular}{|c|c|c|c|}
        \hline
        Classification & Accuracy & Specificity & Sensitivity\\
        \hline
        System level & 98.13\% & 95.12\% & 100\% \\
        \hline
        Single, panel-level & 98.78\% & 97\% & 100\% \\
        \hline
        Multiple panel-level & 97.2\% & 97.06\% & 97.26\% \\
        \hline
    \end{tabular}
    \caption{Classification Metrics}
    \label{class_metrics}
\end{table}

\emph{Results}
Our experiments  demonstrate the efficacy of our fault detection and classification methods for real snow faults  as well as synthetically injected single and concurrent panel-level faults.
Our results show that the random forest classifier is an effective
approach for identifying  both system-wide  faults as well as
faults that occur on a subset of panels.
Our approach is able to  classify  snow, partial occlusion and open circuit faults with accuracy of more than 97\% in terms of overall accuracy, specificity, and sensitivity.

%% file: related.tex
\section{Related Work}
There has been significant work on predicting power output for solar sites~\cite{pvlib, solar-tk, kpv, lorenz2007forecast, lorenz2009irradiance, perez-2018, sharma2011predicting}. All of these studies predict only system level output by using long term historical data for model training~\cite{perez-2018, sharma2011predicting}, small amount of historical data for estimating system parameters~\cite{solar-tk}, system configuration details~\cite{pvlib, lorenz2007forecast, lorenz2009irradiance}, or output from a nearby site~\cite{kpv}. None of the studies predict the individual panel level output, their prediction for all of the panels would be the same. Furthermore, while the anomaly detection and classification is not the key goal, some of these studies can be used to detect panels whose output significantly varies from the system level output. However, a 20-30\% error reported by these approaches in system level output prediction will limit their anomaly detection and classification accuracy. 

There is also significant prior work on anomaly detection and classification in solar photovoltaic systems, that can be broadly classified into model-based approaches~\cite{kang2012diagnosis, hu2013photovoltaic, kim2015photovoltaic, dhimish2017parallel, garoudja2017statistical} and machine learning based~\cite{de2018predictive-anomaly, gao2015s, harrou2019svm1anomaly, pereira2018autoencoder-anomaly, zhao2018hierarchical-anomaly, mekki2016artificial, chine2016novel, liu2017fault, liu2018condition, zhu2018fault} approaches. 
Model based approaches produce accurate analytical results, but require PV module's specifications and cannot adapt to complex PV systems if the pre-defined parameters change with dynamic environment~\cite{liu2017fault}. Some of the studies use power output data from nearby solar sites~\cite{vergura2018test-based-anomaly, iyengar2018solarclique} to detect and classify anomalies.  In \cite{vergura2018test-based-anomaly}, authors compare the performance of different solar arrays at the same site, but do not do anomaly classification. 

To the best of our knowledge, there is no prior work on classifying panel-level anomalies. All of the aforementioned approaches target system-level anomaly detection and are not suitable for panel-level anomaly classification tasks.  We extend the anomaly detection and classification capability to panel level, where we are able to classify various types of faults, i.e. snow, object covering, and electrical faults, on a single or multiple panels.




%% file: conclusion.tex
\section{Conclusions}

In this paper, we proposed SunDown, a sensorless approach to detecting per-panel anomalies in residential solar arrays. Our approach uses a model-driven approach that leverages correlations between the power produced by adjacent panels to detect deviations from expected behavior. SundDown can handle faults in multiple panels and determine the probable cause of anomalies. We evaluated SunDown using two year panel-level generation data from the from a real site and a manually gathered dataset of various faults. Our approach requires data from only 5 panels for accurate prediction, is agnostic to weather characteristics, and yields high accuracy even when panels from different roof geometries are used. We show that our approach is accurate in predicting panel level output with a MAPE of 2.98\% and can correctly classify anomalies with >97\% accuracy. We released the per-panel dataset from the real site and the manually generated dataset of various faults for research use. 